\documentclass[submission,copyright,creativecommons]{eptcs}
\providecommand{\event}{ACL2 2020} 
\usepackage{underscore}           

\title{Generating Mutually Inductive Theorems from Concise Descriptions}
\author{Sol Swords
\institute{Centaur Technology, Inc.}
\email{sswords@centtech.com}
}
\def\titlerunning{Generating Mutually Inductive Theorems}
\def\authorrunning{S. Swords}

\usepackage{amsmath}
\usepackage{listings}
\usepackage{preview}
\usepackage{xstring}
\usepackage{breakurl}

\newcommand{\va}[1]{\ensuremath{\operatorname{\mathit{#1}}}}
\newcommand{\Code}[1]{\texttt{#1}}
\newcommand{\Codeh}[1]{\texttt{\StrSubstitute{#1}{-}{-\allowbreak{}}}}
\newcommand{\Opt}[1]{\textit{[}\hspace{-2pt} #1\hspace{-2pt} \textit{]}}
\PreviewEnvironment{lstlisting}
\PreviewMacro[{{}}]{\Codeh}

\begin{document}
\maketitle

\begin{abstract}
  We describe \texttt{defret-mutual-generate}, a utility for proving
  ACL2 theorems about large mutually recursive cliques of functions.
  This builds on previous tools such as \texttt{defret-mutual} and
  \texttt{make-flag}, which automate parts of the process but still
  require a theorem body to be written out for each function in the
  clique.  For large cliques, this tends to mean that certain common
  hypotheses and conclusions are repeated many times, making proofs
  difficult to read, write, and maintain.  This utility automates
  several of the most common patterns that occur in these forms, such
  as including hypotheses based on formal names or types.  Its input
  language is rich enough to support forms that have some common parts
  and some unique parts per function.  One application of
  \texttt{defret-mutual-generate} has been to support proofs about the
  FGL rewriter, which consists of a mutually recursive clique of 49
  functions.  The use of this utility reduced the size of the forms
  that express theorems about this clique by an order of magnitude. It
  also greatly has reduced the need to edit theorem forms when
  changing definitions in the clique, even when adding or removing
  functions.
\end{abstract}

\section{Introduction}

Mutual recursion is used fairly frequently in ACL2, but it is still
relatively rare to prove significant theorems about mutually recursive
functions.  Most theorems in the ACL2 community books that mention
mutually recursive functions are generated by utilities such as
\Codeh{fty::deftypes}, \Codeh{fty::deffixequiv-mutual}, or the
\Codeh{:returns} feature of \Codeh{std::defines} \cite{acl2:doc}.  We
posit that the reason for this is not that mutually recursive
algorithms are uninteresting, but that perhaps few users know of the
existing tools that support proofs about them.  Another impediment is
that it is usually necessary to write several variations of the
desired theorem, one for each function in the clique, in order to
prove a theorem by mutual induction.

In this paper we first describe existing processes for proving
inductive theorems about mutual recursions, including the utilities
\Codeh{make-flag} and \Codeh{defret-mutual}.  We then describe a new
utility, \Codeh{defret-mutual-generate}, that builds on these and
automates the generation of such theorems, using schemas that address
many common usage patterns.  This utility was developed alongside the
FGL rewriter, the core definitions of which are in a clique of 49
mutually-recursive functions.  We calculate that without the use of
this utility, the forms expressing the core invariants and correctness
theorems about the FGL rewriter would have been an order of magnitude
bigger.  Furthermore, the use of this utility saves the need to edit
all of these theorem forms every time a function in the clique is
added, removed, or its input/output signature changed---usually most
of the theorem forms can be left unmodified.


Many of the utilities described here are more thoroughly documented in
the combined ACL2 and community books manual \cite{acl2:doc}.  We'll
refer simply to ``the manual'' as a shorthand when we reference the
respective documentation topics for such utilities.

\section{Mutually Inductive Proofs}

We first show a simple mutually inductive proof about a mutually
recursive clique of functions, then describe how to scale this proof
strategy to more complicated functions.  For our example, we'll
prove a theorem about a term substitution algorithm,
\Codeh{subst-term}, defined in Listing \ref{subst term def}.

\lstset{basicstyle=\footnotesize \ttfamily, language=Lisp}

\begin{lstlisting}[float, label=subst term def, caption={Definitions of \Code{subst-term}, \Code{ev-term}, and \Code{ev-alist}}]
(mutual-recursion
 (defun subst-term (x alist)
   (cond ((not x) nil)
         ((symbolp x) ;; variable
          (cdr (assoc-equal x alist)))
         ((atom x) nil) ;; malformed
         ((eq (car x) 'quote) x)
         (t ;; function or lambda call
          (cons (car x)
                (subst-termlist (cdr x) alist)))))
 (defun subst-termlist (x alist)
   (if (atom x)
       nil
     (cons (subst-term (car x) alist)
           (subst-termlist (cdr x) alist)))))

(defevaluator ev-term ev-termlist nil :namedp t)

(defun ev-alist (x env)
  (if (atom x)
      nil
    (cons (cons (caar x) (ev-term (cdar x) env))
          (ev-alist (cdr x) env))))
\end{lstlisting}

The theorem we will prove states its semantics with respect to
\Codeh{ev-term}, a standard term evaluator created with
\Codeh{defevaluator}~\cite{05-hunt-meta}.  The theorem we want is shown
in Listing \ref{ev of subst thm}.

\begin{lstlisting}[float, label=ev of subst thm, caption={Evaluation of \Code{subst-term} theorem}]
(defthm ev-term-of-subst-term
  (equal (ev-term (subst-term x alist) env)
         (ev-term x (ev-alist alist env))))
\end{lstlisting}

\begin{lstlisting}[float, label=ev of subst list thm, caption={Evaluation of \Code{subst-termlist} theorem}]
(defthm ev-termlist-of-subst-termlist
  (equal (ev-termlist (subst-termlist x alist) env)
         (ev-termlist x (ev-alist alist env))))
\end{lstlisting}

The problem we'll encounter if we try to prove
\Codeh{ev-term-of-subst-term} is that we need a lemma,
\Codeh{ev-termlist-of-subst-termlist} (Listing~\ref{ev of subst list
  thm}).  But we can't prove that lemma by itself, because we need the
original \Codeh{ev-term-of-subst-term}---that is, we need to prove the
two theorems via mutual induction.  The simplest way to prove these
two theorems is to prove their conjunction,
\Codeh{ev-term/list-of-subst-term/list} (Listing~\ref{ev of subst
  mutual thm}), by an induction scheme that recurs on the \Codeh{car}
and \Codeh{cdr} of \Codeh{x}.

\begin{lstlisting}[float, label=ev of subst mutual thm, caption={Mutually-inductive evaluation theorem}]
(defun subst-term-ind (x)
  (and (consp x)
       (list (subst-term-ind (car x))
             (subst-term-ind (cdr x)))))

(defthm ev-term/list-of-subst-term/list
  (and (equal (ev-term (subst-term x alist) env)
              (ev-term x (ev-alist alist env)))
       (equal (ev-termlist (subst-termlist x alist) env)
              (ev-termlist x (ev-alist alist env))))
  :hints (("goal" :induct (subst-term-ind x))))
\end{lstlisting}

In this approach to the problem, we prove the conjunction of the
mutually inductive theorems using a custom induction scheme, which
typically must match the recursive calls of all the functions of the clique.  
Here \Codeh{subst-term-ind} suffices because it recurs on
\Codeh{(cdr x)} when \Codeh{x} is a function or lambda call term, as in
\Codeh{subst-term}, and it recurs on both \Codeh{(car x)} and
\Codeh{(cdr x)} when \Codeh{x} is a cons, as does \Codeh{subst-termlist}.

There are two problems with applying this approach to larger problems.
First, it isn't always easy to hand-craft an induction scheme that
contains a superset of all the recursive calls of a clique.  Second, these
sorts of induction schemes will produce too many induction hypotheses.
In this example, we still have a fast proof despite generating several 
useless induction hypotheses.  But for larger
cliques, the number of induction hypotheses will usually grow as the
number of functions in the clique times the number of different
recursive calls, which can quickly overwhelm the prover with useless
hypotheses.

These two problems can be addressed by instead doing the induction
using a \textit{flag function} version of the mutual recursion.  Any
mutually-recursive clique of functions can be transformed into a
single function whose formals are the union of the formals of the
clique functions along with an extra formal called the flag, which
tells which function of the clique the flag function should emulate.
This technique dates back at least to 1984, when Boyer and Moore
\cite{boyer1984mechanical} noted:
\begin{quote}
  ...it is well known that mutual recursion can be eliminated by the
  trick of defining a single function that has an extra ``flag''
  argument...
\end{quote}
The flag function can be proved equal to the functions of the
original mutual recursion, dispatched by the flag.  A flag function for
\Codeh{subst-term} and its equivalence theorem is shown in
Listing~\ref{subst term flag fn}.

\begin{lstlisting}[float, label=subst term flag fn, caption={Flag function for \Code{subst-term}}]
(defun subst-term-flag (flag x alist)
  (case flag
    (subst-term
     (cond ((not x) nil)
           ((symbolp x) ;; variable
            (cdr (assoc-equal x alist)))
           ((atom x) nil) ;; malformed
           ((eq (car x) 'quote) x)
           (t ;; function or lambda call
            (cons (car x)
                  (subst-term-flag 'subst-termlist (cdr x) alist)))))
    (t ;; subst-termlist
     (if (atom x)
         nil
       (cons (subst-term-flag 'subst-term (car x) alist)
             (subst-term-flag 'subst-termlist (cdr x) alist))))))

(defthm subst-term-flag-equals-subst-term
  (equal (subst-term-flag flag x alist)
         (case flag
           (subst-term
            (subst-term x alist))
           (t
            (subst-termlist x alist)))))
\end{lstlisting}

The flag function can then be used as an induction scheme to prove
mutually-inductive theorems about the original functions, using the
flag variable to distinguish between the cases.  That is, instead of
proving the conjunction of all the mutually inductive theorems, we
prove that each of them is true when the flag is the corresponding
value, as shown in Listing~\ref{subst term flag theorem}.  This form
of the theorem usually doesn't produce good rewrite or
other rule classes because of the presence of the extra flag variable.
But instantiating this theorem with the various values of the flag
variable is an easy way to prove the original mutually-inductive
theorems.

\begin{lstlisting}[float, label=subst term flag theorem,
  caption={Flag-style proof of \Code{subst-term} evaluation}]
(defthm ev-term/list-of-subst-term/list-lemma
  (case flag
    (subst-term (equal (ev-term (subst-term x alist) env)
                       (ev-term x (ev-alist alist env))))
    (t ;; subst-termlist
     (equal (ev-termlist (subst-termlist x alist) env)
            (ev-termlist x (ev-alist alist env)))))
  :hints (("goal" :induct (subst-term-flag flag x alist)))
  :rule-classes nil)

(defthm ev-term-of-subst-term
  (equal (ev-term (subst-term x alist) env)
         (ev-term x (ev-alist alist env)))
  :hints (("goal" :use ((:instance ev-term/list-of-subst-term/list-lemma
                         (flag 'subst-term))))))

(defthm ev-termlist-of-subst-termlist
  (equal (ev-termlist (subst-termlist x alist) env)
         (ev-termlist x (ev-alist alist env)))
  :hints (("goal" :use ((:instance ev-term/list-of-subst-term/list-lemma
                         (flag 'subst-termlist))))))
\end{lstlisting}

There are several advantages of this scheme over the previous approach
of proving the conjunction using a custom induction scheme.  It is
easy to automate because the transformation of the clique to a flag
function is straightforward.  The induction scheme is specific to each
function of the clique, so that after some simple case splitting, only
the particular induction hypotheses needed for a given case are left.
Because the flag function emulates the original clique, its induction
scheme even works when there are \textit{reflexive} recursive calls,
that is, recursive calls on the results of other recursive calls.

\section{Flag function method using \texttt{make-flag}}

The macro \Codeh{make-flag} automates the flag function method shown in
the previous section.  A reference for the full feature set of
\Codeh{make-flag} is in the manual~\cite{acl2:doc} and
beyond the scope of this paper, but we briefly describe what it does
by example.

The events of Listing~\ref{make flag example} show how to prove the
two mutually-inductive theorems of the previous section.  The
\Codeh{make-flag} event admits a flag function and equivalence theorem,
similar to the hand-coded events of Listing~\ref{subst term flag fn}.
It also defines a new macro named \Codeh{defthm-subst-term-flag} that
uses the flag function to prove a mutually-inductive set of theorems
about the original clique.  We'll call this sort of macro a
\textit{flag defthm macro}. For each theorem, the user must specify
which function of the clique (and therefore which value of the flag)
it corresponds to.  It generates an \Codeh{encapsulate} event that
contains essentially the events of Listing~\ref{subst term flag
  theorem}, with the original lemma local to the encapsulate but the
other two theorems exported.

\begin{lstlisting}[float, label=make flag example,
  caption={Proof of \Code{subst-term} evaluation using \Code{make-flag}}]
(flag::make-flag subst-term-flag subst-term)

(defthm-subst-term-flag
  (defthm ev-term-of-subst-term
    (equal (ev-term (subst-term x alist) env)
           (ev-term x (ev-alist alist env)))
    :flag subst-term)
  (defthm ev-termlist-of-subst-termlist
    (equal (ev-termlist (subst-termlist x alist) env)
           (ev-termlist x (ev-alist alist env)))
    :flag subst-termlist))
\end{lstlisting}

\subsection{Using \texttt{defun-sk} Instead of Specialized Induction Schemes}

In some cases a proof seems to require an induction scheme that isn't
exactly the one generated by the main (mutually) recursive function
involved.  For example, in Listing~\ref{remove return last} we show
the mutually-recursive definitions of \Codeh{remove-return-last-term}
and \Codeh{remove-return-last-termlist} and a pair of theorems about
the clique (where \Codeh{rl-ev} and \Codeh{rl-ev-list} are a term/list
evaluator pair).  Intuitively we might expect to prove these using the
induction scheme of a flag function generated from the clique.
However, this induction doesn't suffice to prove these theorems
directly, because the induction hypothesis we need for the
\Codeh{lambda} case has a substitution for \Codeh{env} as well as for
\Codeh{x}.

\begin{lstlisting}[float, label=remove return last,
  caption={Definition and desired theorem about \Code{remove-return-last-term}},
  escapechar=^]
(mutual-recursion
 (defun remove-return-last-term (x)
   (cond ((atom x) x)
         ((eq (car x) 'quote) x)
         ((eq (car x) 'return-last)
          (remove-return-last-term (cadddr x)))
         ((consp (car x))
          ;;lambda
          ^\`{}^((lambda ,(cadar x)
              ,(remove-return-last-term (caddar x)))
             . ,(remove-return-last-termlist (cdr x))))
         (t (cons (car x) (remove-return-last-termlist (cdr x))))))
 (defun remove-return-last-termlist (x)
   (if (atom x)
       nil
     (cons (remove-return-last-term (car x))
           (remove-return-last-termlist (cdr x))))))

(defevaluator rl-ev rl-ev-list ((return-last x y z)) :namedp t)

(defthm remove-return-last-term-correct
  (equal (rl-ev (remove-return-last-term x) env)
         (rl-ev x env)))

(defthm remove-return-last-termlist-correct
  (equal (rl-ev-list (remove-return-last-termlist x) env)
         (rl-ev-list x env)))
\end{lstlisting}
         
One way to solve this problem is to write a custom induction scheme
that produces the same substitutions for \Codeh{x} as in the flag
function but takes \Codeh{env} as an additional input and passes the
correct substitution for that \Codeh{env} in the lambda case.  This
could even be written as a second mutual recursion for which a second
flag function is automatically generated.  However, for more
complicated cliques of functions, introducing a second similar mutual
recursion is rather unwieldy and violates the Don't Repeat Yourself
(DRY) principle; the two copies of the mutual recursion must always be
maintained in parallel, and it wouldn't be easy in general to remove
this duplicated code by generating both versions from one codebase.

An alternative is to use the flag induction of the original mutual
recursion but with universal quantification of the variables that need
specialized substitutions in some induction hypotheses.  That is,
instead of proving \Codeh{(p x env)} by induction, we prove
\Codeh{(forall env (p x env))} by induction.  This sounds strange in
the ACL2 logic where all free variables of a theorem are implicitly
universally quantified.  However, when we induct on the latter using
an induction scheme that applies a substitution to \Codeh{x}, we get to
assume \Code{(forall env (p $\sigma(\Code{x})$ env))} instead of
\Code{(p $\sigma(\Code{x})$ env)}.  To do this we introduce a
quantifier function using \Codeh{defun-sk} for each of the mutually
inductive theorems, prove the quantifier functions true via the flag
induction, and then prove the original theorems we wanted as
corollaries of those lemmas.  We show the process in
Listing~\ref{return last proof}, eliding the termlist versions of the
\Codeh{defun-sk}, lemma, and final theorem.  Of course, there is some
lack of DRYness in this method as well, but repeating the theorem
bodies is likely preferable to repeating the function definitions,
and it would be easier to use macros to streamline this method as well.

\begin{lstlisting}[float, label=return last proof,
  caption={Theorem about \Code{remove-return-last-term} proved by induction over quantification},
  escapechar=^]
  (defun-sk remove-return-last-term-correct-cond (x)
    (forall env
            (equal (rl-ev (remove-return-last-term x) env)
                   (rl-ev x env)))
    :rewrite :direct)

  (defun-sk remove-return-last-termlist-correct-cond (x) ...)

  (defthm-remove-return-last-flag
    (defthm remove-return-last-term-correct-lemma
      (remove-return-last-term-correct-cond x)
      :hints ((and stable-under-simplificationp
                   ^\`{}^(:expand (,(car (last clause))))))
      :flag remove-return-last-term
      :rule-classes nil)
     ...)

  (defthm remove-return-last-term-correct
    (equal (rl-ev (remove-return-last-term x) env)
           (rl-ev x env))
    :hints (("goal" :use remove-return-last-term-correct-lemma)))
\end{lstlisting}

Note the stable-under-simplification hint \Codeh{\`{}(:expand (,(car (last
  clause))))} in the inductive lemma.  This is often a useful hint for
these proofs because it opens the occurrence of the Skolemized
function in the conclusion, but not the inductive hypotheses.  To
prove a universal quantifier introduced with \Codeh{defun-sk} true, we
want to expand it and prove that its body is true of the witness,
whereas if we are assuming it true it is more convenient to leave it
unexpanded so that its rewrite rule may be applied.

\section{Proofs using \texttt{defines} and \texttt{defret-mutual}}

The utilities \Codeh{define} and \Codeh{defines} add several features to
(respectively) \Codeh{defun} and \Codeh{mutual-recursion}.  Their full
documentation is in the manual~\cite{acl2:doc}, and we will touch on
only a few salient features.

The main advantage to \Codeh{define} and \Codeh{defines} that we exploit
to generate mutually inductive theorems is that they store extra data
in a table about the functions and mutual recursions they generate.
In particular, they allow the return values of functions to be named,
and provide a syntax for declaring the types of both return values and
formals.  This type data is important for
\Codeh{defret-mutual-generate}, discussed in the next section.
But simply naming the return values, especially for functions
that return multiple values, allows theorems about such functions to
be written much more concisely.  The \Codeh{defret} utility produces a
\Codeh{defthm} form that binds the return values to the call of the
function on its formals, for the last function defined with
\Codeh{define} by default.  It also supports various other
abbreviations; see the manual~\cite{acl2:doc} for details. For mutual
inductions, \Codeh{defret-mutual} expands to the flag defthm macro of
the mutual recursion most recently introduced with \Codeh{defines},
which by default produces an implicit \Codeh{make-flag} event.

For our \Codeh{subst-term} example, if we recode the function using
\Codeh{defines} then we can do the same proofs in a
\Codeh{defret-mutual} form as shown in Listing~\ref{subst term
  defines}.  Note that because the \Codeh{define} forms for each of the
functions include a \Codeh{:returns} form naming the output of the
function \Codeh{subst}, the variable \Codeh{subst} in the \Codeh{defret}
forms is implicitly bound to the call of the respective functions.

\begin{lstlisting}[float, label=subst term defines,
  caption={Defines form for \Code{subst-term} and evaluation theorem}]
(defines subst-term
  (define subst-term ((x pseudo-termp) (alist pseudo-term-substp))
    :returns (subst)
    (cond ((not x) nil)
          ((symbolp x) ;; variable
           (cdr (assoc-equal x alist)))
          ((atom x) nil) ;; malformed
          ((eq (car x) 'quote) x)
          (t ;; function or lambda call
           (cons (car x)
                 (subst-termlist (cdr x) alist)))))
  (define subst-termlist ((x pseudo-term-listp) (alist pseudo-term-substp))
    :returns (subst)
    (if (atom x)
        nil
      (cons (subst-term (car x) alist)
            (subst-termlist (cdr x) alist))))
  ///

  (defret-mutual ev-term-of-subst-term
    (defret ev-term-of-subst-term
      (equal (ev-term subst env)
             (ev-term x (ev-alist alist env)))
      :fn subst-term)
    (defret ev-termlist-of-subst-termlist
      (equal (ev-termlist subst env)
             (ev-termlist x (ev-alist alist env)))
      :fn subst-termlist)))
\end{lstlisting}
    
\section{Automation using \texttt{defret-mutual-generate}}

For proofs about large mutually-recursive cliques, one of the major
problems is the usual need to include one theorem per function in
order to achieve the correct mutual induction.  In proofs about the
FGL rewriter \cite{fgl-github}, the mutually recursive clique in
question contains 49 functions, all of which take and return two
stobjs, \Codeh{interp-st} and \Codeh{state}, and most of which have one
or two additional arguments and return values.  To prove even simple
invariants would require writing flag defthm macro forms of well over
300 lines or \Codeh{defret-mutual} forms of well over 100
lines.

Previous projects in the ACL2 community books \cite{acl2:doc} that
also ran into this problem have used ad hoc solutions such as
custom-built macros to support proofs.  For two examples, see the GL
interpreter, whose proofs are supported by the macro
\Codeh{def-glcp-interp-thm}, and the VL expression and statement
parsers, which use custom theorem generator functions such as
\Codeh{vl-val-when-error-claim}, \Codeh{vl-warning-claim}, etc.  FGL's
mutually inductive proofs were instead supported by a more
general-purpose utility, \Codeh{defret-mutual-generate}.  There are 22
sets of theorems about the FGL rewriter, all generated by this
utility.  Of those, 17 are mutual inductions and the other five are
corollaries of mutually-inductive theorems for which induction isn't
needed.  The average size of these \Codeh{defret-mutual-generate} forms
is 41 lines.  This average is dominated by the final correctness
theorem, which is larger because most functions in the mutual
recursion need unique correctness statements; this form is 430 lines
long, and the average omitting this one is 23 lines.

As a simple example, we show the first \Codeh{defret-mutual-generate}
form in Listing~\ref{scratch isomorphic thm}.  The theorem bodies are
generated from the \Codeh{:return-concls} argument, which essentially
says ``for each function that has a return value named
\Codeh{new-interp-st}, prove the following conclusion.''  In
Listing~\ref{scratch isomorphic expansions} we show two steps of the
expansion of the form, heavily elided.  First it expands to a
\Codeh{defret-mutual} form containing 49 \Codeh{defret} forms.  This then
expands, mainly by adding the \Code{b*} bindings of the return
values for each function, to a \Codeh{defthm-fgl-interp-flag}
containing 49 \Codeh{defthm} forms.  In both cases we have omitted all
but two of the 49 forms from the listing.

\begin{lstlisting}[float, label=scratch isomorphic thm,
  caption={Simple \Code{defret-mutual-generate} form}]
(defret-mutual-generate interp-st-scratch-isomorphic-of-<fn>
  :return-concls ((new-interp-st
                   (interp-st-scratch-isomorphic new-interp-st
                                                 (double-rewrite interp-st))))
  :hints ((fgl-interp-default-hint 'fgl-interp-term id nil world))
  :mutual-recursion fgl-interp)
\end{lstlisting}

\begin{lstlisting}[float, label=scratch isomorphic expansions,
  caption={Expansions of a \Code{defret-mutual-generate} form}]
(defret-mutual interp-st-scratch-isomorphic-of-<fn>
  (defret interp-st-scratch-isomorphic-of-<fn>
    (interp-st-scratch-isomorphic new-interp-st (double-rewrite interp-st))
    :fn fgl-interp-test)
  ...
  (defret interp-st-scratch-isomorphic-of-<fn>
    (interp-st-scratch-isomorphic new-interp-st (double-rewrite interp-st))
    :fn fgl-interp-merge-branch-args)
  :mutual-recursion fgl-interp)

(defthm-fgl-interp-flag interp-st-scratch-isomorphic-of-<fn>
  (defthm interp-st-scratch-isomorphic-of-fgl-interp-test
    (b* (((mv ?xbfr ?new-interp-st ?new-state)
          (fgl-interp-test x interp-st state)))
      (interp-st-scratch-isomorphic new-interp-st (double-rewrite interp-st)))
    :flag fgl-interp-test)
  ...
  (defthm interp-st-scratch-isomorphic-of-fgl-interp-merge-branch-args
    (b* (((mv acl2::?args ?new-interp-st ?new-state)
          (fgl-interp-merge-branch-args testbfr
                                        thenargs elseargs interp-st state)))
      (interp-st-scratch-isomorphic new-interp-st (double-rewrite interp-st)))
    :flag fgl-interp-merge-branch-args))
\end{lstlisting}

A more complicated example is shown in Listing~\ref{bfrs ok thm}.
This form generates a \Codeh{defret-mutual} that is 400 lines long, and
unlike the previous example the theorems generated aren't all the
same.  The theorem bodies are generated by applying a set of rules to
the function signatures, namely the \Codeh{define} formals and returns.
These rules are determined by the arguments to the
\Codeh{defret-mutual-generate} form.

\begin{lstlisting}[float, label=bfrs ok thm,
  caption={Expansions of a \Code{defret-mutual-generate} form},
  escapechar=^]
(defret-mutual-generate interp-st-bfrs-ok-of-<fn>
  :formal-hyps
  (((interp-st-bfr-p x)           (lbfr-p x))
   ((fgl-object-p x)              (lbfr-listp (fgl-object-bfrlist x)))
   ((fgl-objectlist-p x)          (lbfr-listp (fgl-objectlist-bfrlist x)))
   ((fgl-object-bindings-p x)     (lbfr-listp (fgl-object-bindings-bfrlist x)))
   (interp-st                     (interp-st-bfrs-ok interp-st))
   ((constraint-instancelist-p x) (lbfr-listp
                                   (constraint-instancelist-bfrlist x))))
  :return-concls
  ((xbfr                          (lbfr-p xbfr new-logicman))
   ((fgl-object-p x)              (lbfr-listp (fgl-object-bfrlist x)
                                              new-logicman))
   ((fgl-objectlist-p x)          (lbfr-listp (fgl-objectlist-bfrlist x)
                                              new-logicman))
   (new-interp-st                 (interp-st-bfrs-ok new-interp-st)))
  :rules
  ((t (:add-bindings ((?logicman (interp-st->logicman interp-st))
                      (?new-logicman (interp-st->logicman new-interp-st)))))
   ((or (:fnname fgl-rewrite-try-rules)
        (:fnname fgl-rewrite-try-rule)
        (:fnname fgl-rewrite-try-rewrite)
        (:fnname fgl-rewrite-try-meta)
        (:fnname fgl-rewrite-binder-try-rules)
        (:fnname fgl-rewrite-binder-try-rule)
        (:fnname fgl-rewrite-binder-try-rewrite)
        (:fnname fgl-rewrite-binder-try-meta)
        (:fnname fgl-rewrite-try-rules3))
    (:add-hyp (scratchobj-case
               (stack^\$^a-top-scratch
                (double-rewrite (interp-st->stack interp-st)))
               :fgl-objlist))))
  :hints ((fgl-interp-default-hint 'fgl-interp-term id nil world)
          '(:do-not-induct t))
  :mutual-recursion fgl-interp)
\end{lstlisting}

Some of the rules set up by this form could be read in English as follows:
\begin{itemize}
\item For each function of the clique that has a formal declared to be
  type \Codeh{interp-st-bfr-p}, add a hypothesis \Codeh{(lbfr-p x)},
  where \Codeh{x} is the formal name, to the theorem for that function.
\item For each function that has a formal named \Codeh{interp-st}, add a hypothesis
  \Codeh{(interp-st-bfrs-ok interp-st)}.
\item For each function that has a return value named \Codeh{xbfr}, add a conclusion
  \Codeh{(lbfr-p xbfr new-logicman)}.
\item For each function that has a return value declared to be type
  \Codeh{fgl-object-p}, add a conclusion
  \Codeh{(lbfr-listp (fgl-object-bfrlist x) new-logicman)},
  where \Codeh{x} is the return name.
\item To every function's theorem, add the \Codeh{B*} bindings for
  \Codeh{logicman} and \Codeh{new-logicman} as listed (see the \Codeh{B*}
  topic in the manual~\cite{acl2:doc}).
\item For every function in the list \Codeh{fgl-rewrite-try-rules},
  etc., add the given \Codeh{scratchobj-case} hypothesis.
\end{itemize}

\subsection{Operation of \texttt{defret-mutual-generate}}

\Codeh{Defret-mutual-generate} produces a \Codeh{defret-mutual} form by
applying a set of rules to each function in a mutually recursive
clique.  These rules may be given directly as arguments to
\Codeh{defret-mutual-generate}, but may also be produced by
abbreviations such as \Codeh{:formal-hyps} and \Codeh{:return-concls}, described below.

When applying the rules to each function, each rule has a
\textit{condition} under which it will take effect and a list of
\textit{actions} that update a structure from which a \Codeh{defret}
form may be generated.  This structure contains the following fields:
\begin{itemize}
\item \textit{Theorem name.}
\item \textit{Top hyps.} A list of top-level hypotheses (implicitly conjoined), which
  apply to the whole conclusion.
\item \textit{Hyp/conclusion stack.} An ordered list containing conclusions (implicitly conjoined) as
  well as \textit{push-hyp} and \textit{pop-hyp} entries; each
  hypothesis added by a push-hyp entry affects the conclusions
  listed subsequently until the corresponding occurrence of pop-hyp.
\item \textit{Bindings.} An ordered list of \Codeh{B*} bindings to be
  applied to all hypotheses and conclusions generated.
\item \textit{Keywords.} Keyword/value arguments such as \Codeh{:hints} and \Codeh{:rule-classes}.
\end{itemize}
Initially, there are no hypotheses, conclusions, bindings, or keywords in this
structure.  Rules may add/\allowbreak{}push/\allowbreak{}pop hypotheses and add conclusions, add
bindings, change the theorem name, and add keyword arguments.  When
all the rules have been applied, a \Codeh{defret} form is generated
from the final structure unless the structure contains no conclusion,
in which case it is skipped.

The conditions governing the rules may be a Boolean AND/OR/NOT
combination of the following primitive expressions, along with \Codeh{t} and \Codeh{nil}:
\begin{itemize}
  \item \Code{(:fnname \va{name})} checks that the name of the function is $\va{name}$.
  \item \Code{(:has-formal \Opt{:name \va{name}} \Opt{:type \va{type}})} checks
    that the function has a formal satisfying the listed criteria. The
    name and type options may be used individually or in combination.
  \item \Code{(:has-return \Opt{:name \va{name}} \Opt{:type \va{type}})} checks that the
    function has a return value satisfying the listed criteria.
\end{itemize}

The actions may be any of the following:
\begin{itemize}
\item \Code{(:add-hyp \va{term})} adds \va{term} as a top-level hypothesis.
\item \Code{(:add-concl \va{term})} adds \va{term} as a conclusion to the hyp/conclusion stack.
\item \Code{(:add-bindings \va{bindings})} appends \va{bindings} to the end of the current bindings list.
\item \Code{(:push-hyp \va{term})} adds a push-hyp entry \va{term} to the hyp/conclusion stack.
\item \Code{(:pop-hyp)} adds a pop-hyp entry to the hyp/conclusion
  stack, cancelling the effect of the previous push-hyp event on
  subsequently added conclusions.
\item \Code{(:each-formal :type \va{type} :var \va{var} :action
    \va{action})}, where \va{action} is either an \Codeh{:add-hyp},
  \Codeh{:push-hyp}, \Codeh{:pop-hyp}, or \Codeh{:add-concl} form, does
  the given action once for each formal of the given type,
  substituting the formal name in each case for \va{var} in the added
  hyp/conclusion term.
\item \Code{(:each-return :type \va{type} :var \va{var} :action \va{action})}
  is similar to \Codeh{:each-formal} but runs instead on each return value.
\item \Code{(:add-keyword \va{key} \va{val})} adds the given keyword/value pair to the keywords.
\item \Code{(:set-thmname \va{template})} sets the theorem name to the
  given template symbol, where any substring \Codeh{<FN>} of the
  template is replaced by the name of the function.
\end{itemize}

The following keywords generate additional rules from the arguments provided:

\begin{itemize}
\item \Code{:formal-hyps} generates hypotheses based on the names or
  types of formals. It takes an argument which is a list of elements
  of the following two forms:
  \begin{itemize}
  \item \Code{(\va{name} \va{term} \Opt{:type \va{type}})} adds the
    given term as a top-level hypothesis to the theorem of any
    function with a formal of the given name.  If a type is provided,
    it will only be added if that formal is of the given type.  This
    translates to a rule with a \Codeh{:has-formal} condition and an
    \Codeh{:add-hyp} action.
  \item \Code{((\va{type} \va{name}) \va{term})} adds the given term
    as a top-level hypothesis for every formal of the given type,
    binding that formal to \va{name}.  This translates to an
    \Codeh{:each-formal} \Codeh{:add-hyp} rule under condition \Codeh{t}.
  \end{itemize}
\item \Code{:return-concls} is analogous to \Codeh{:formal-hyps},
  generating conclusions based on the names or types of return values.
  The same forms of argument are accepted.
\item \Code{:function-keys} adds keywords to the theorems
  corresponding to function names.  It accepts an argument which is a
  list of entries \Code{(\va{fnname} \va{key} \va{val} \ldots)}.
\end{itemize}

\section{Conclusion}

The utilities described in this paper are effective in reducing the
boilerplate and code duplication that is otherwise necessary for
proving mutually inductive theorems.  In developing the FGL rewriter,
which is a 49-function mutually recursive clique, these tools were
used to great effect in proving the necessary theorems, including its
semantic correctness with respect to an evaluator.  We have made many
revisions to the FGL rewriter since its correctness proofs were first
completed, including adding and removing several functions from the
mutual recursion.  However, because we use
\Codeh{defret-mutual-generate} to produce the theorems about this
mutual recursion, we usually find that the only one that needs to be
significantly updated is the final semantic correctness theorem.  We
have also found it to be advantageous to split functions from
the FGL rewriter into smaller mutually-recursive parts, since this
makes each step of the inductive proof smaller.  Normally, this would
mean that proof scripts would need to grow larger in order to
accommodate the new functions of the clique, but again we find that
most of the \Code{defret-mutual-generate} forms that generate our
proofs need no modification.

\bibliographystyle{eptcs}
\bibliography{../bib.bib}

\begin{thebibliography}{1}
\providecommand{\bibitemdeclare}[2]{}
\providecommand{\surnamestart}{}
\providecommand{\surnameend}{}
\providecommand{\urlprefix}{Available at }
\providecommand{\url}[1]{\texttt{#1}}
\providecommand{\href}[2]{\texttt{#2}}
\providecommand{\urlalt}[2]{\href{#1}{#2}}
\providecommand{\doi}[1]{doi:\urlalt{http://dx.doi.org/#1}{#1}}
\providecommand{\bibinfo}[2]{#2}

\bibitemdeclare{misc}{acl2:doc}
\bibitem{acl2:doc}
\bibinfo{author}{\surnamestart {ACL2 Community}\surnameend}
  (\bibinfo{year}{Accessed: 2020}): \emph{\bibinfo{title}{{ACL2+Books}
  Documentation}}.
\newblock
  \urlprefix\url{http://www.cs.utexas.edu/users/moore/acl2/manuals/current/manual/index.html}.

\bibitemdeclare{article}{boyer1984mechanical}
\bibitem{boyer1984mechanical}
\bibinfo{author}{Robert~S. \surnamestart Boyer\surnameend} \&
  \bibinfo{author}{J~Strother \surnamestart Moore\surnameend}
  (\bibinfo{year}{1984}): \emph{\bibinfo{title}{A mechanical proof of the
  unsolvability of the halting problem}}.
\newblock {\sl \bibinfo{journal}{Journal of the ACM (JACM)}}
  \bibinfo{volume}{31}(\bibinfo{number}{3}), pp. \bibinfo{pages}{441--458},
  \doi{10.1145/828.1882}.

\bibitemdeclare{inproceedings}{05-hunt-meta}
\bibitem{05-hunt-meta}
\bibinfo{author}{Warren~A. \surnamestart Hunt~Jr.\surnameend},
  \bibinfo{author}{Matt \surnamestart Kaufmann\surnameend},
  \bibinfo{author}{Robert~Bellarmine \surnamestart Krug\surnameend},
  \bibinfo{author}{J.~Strother \surnamestart Moore\surnameend} \&
  \bibinfo{author}{Eric~Whitman \surnamestart Smith\surnameend}
  (\bibinfo{year}{2005}): \emph{\bibinfo{title}{Meta Reasoning in {ACL2}}}.
\newblock In \bibinfo{editor}{Joe \surnamestart Hurd\surnameend} \&
  \bibinfo{editor}{Tom \surnamestart Melham\surnameend}, editors: {\sl
  \bibinfo{booktitle}{Theorem Proving in Higher Order Logics}},
  \bibinfo{publisher}{Springer Berlin Heidelberg}, \bibinfo{address}{Berlin,
  Heidelberg}, pp. \bibinfo{pages}{163--178}, \doi{10.1007/11541868_11}.

\bibitemdeclare{misc}{fgl-github}
\bibitem{fgl-github}
\bibinfo{author}{Sol \surnamestart Swords\surnameend} (\bibinfo{year}{Accessed:
  2020}): \emph{\bibinfo{title}{{FGL} source distribution}}.
\newblock
  \urlprefix\url{https://github.com/acl2/acl2/tree/master/books/centaur/fgl}.

\end{thebibliography}

\end{document}